%% file: main.tex
\PassOptionsToPackage{dvipsnames, table}{xcolor}
\documentclass[conference]{IEEEtran}
\IEEEoverridecommandlockouts

\usepackage{cite}
\usepackage{amsmath,amssymb,amsfonts}
\usepackage{algorithmic}
\usepackage{graphicx}
\usepackage{textcomp}
\def\BibTeX{{\rm B\kern-.05em{\sc i\kern-.025em b}\kern-.08em
    T\kern-.1667em\lower.7ex\hbox{E}\kern-.125emX}}

\usepackage{color,xcolor}
\usepackage[table]{xcolor}
\usepackage{booktabs}
\usepackage{xspace}
\usepackage{newtxmath}
\usepackage{balance}
\usepackage{rotating}
\usepackage{multirow}
\usepackage{url}
\usepackage[colorlinks,citecolor=citecolor, linkcolor=linkcolor, bookmarks=false]{hyperref}
\definecolor{citecolor}{HTML}{1F801F}
\definecolor{linkcolor}{HTML}{ED1C24}
\newcommand{\std}[1]{\scriptsize{#1}}
\newcommand{\method}{DepMamba\xspace}
\newcommand{\etal}{\textit{et al}.\xspace}
\newcommand{\ie}{\textit{i}.\textit{e}.,\xspace}

\definecolor{maroon}{HTML}{800000}
\newcommand{\best}[1]{\textcolor{maroon}{\textbf{#1}}}
\begin{document}
\title{DepMamba\hspace{2.pt}\includegraphics[width=0.24in]{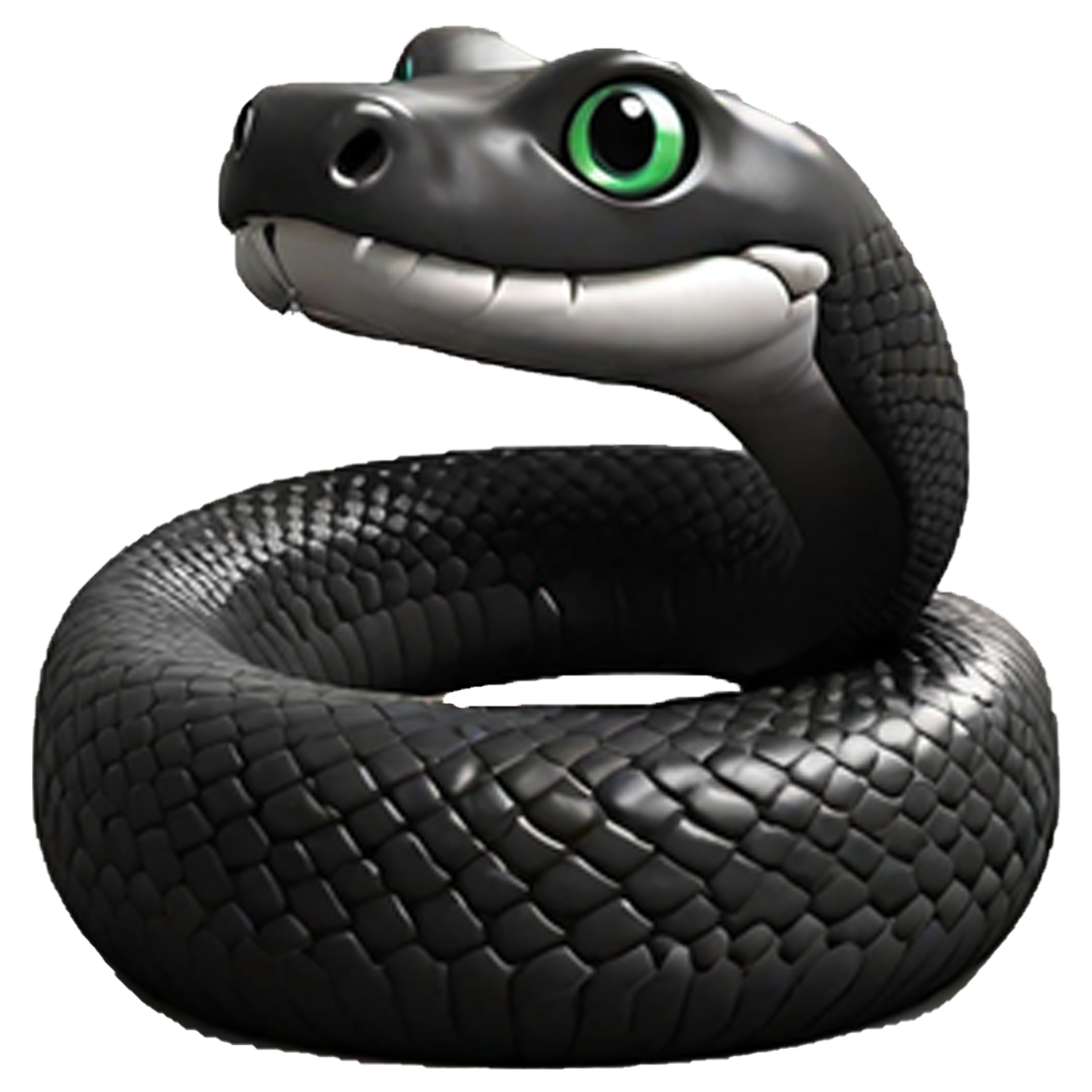}: Progressive Fusion Mamba for Mutilmodal Depression Detection}

\author{
Jiaxin Ye$^1$, Junping Zhang$^2$, Hongming Shan$^{1,\dagger}$\thanks{$\dagger$: Corresponding author.}\\
$^1$ Institute of Science and Technology for Brain-inspired Intelligence, Fudan University, Shanghai, China\\
$^2$School of Computer Science, Fudan University, Shanghai, China \\
jxye22@m.fudan.edu.cn, \{jpzhang, hmshan\}@fudan.edu.cn
}

\maketitle

\input{Section/0_abs}
\input{Section/1_intro}

\input{Section/2_method}
\input{Section/3_exp}
\input{Section/4_summary}

\clearpage
\newpage
{
\bibliographystyle{IEEEbib}

}
\end{document}

%% file: Section/0_abs.tex
\begin{abstract}
Depression is a common mental disorder that affects millions of people worldwide. 
Although promising, current multimodal methods hinge on aligned or aggregated multimodal fusion, suffering two significant limitations: (\textbf{i}) \emph{inefficient} long-range temporal modeling, and (\textbf{ii}) \emph{sub-optimal} multimodal fusion between intermodal fusion and intramodal processing.
In this paper, we propose an audio-visual progressive fusion Mamba for multimodal depression detection, termed DepMamba. 
DepMamba features two core designs: hierarchical contextual modeling and progressive multimodal fusion. 
On the one hand, hierarchical modeling introduces convolution neural networks and Mamba to extract the local-to-global features within long-range sequences. 
On the other hand, the progressive fusion first presents a multimodal collaborative State Space Model (SSM) extracting intermodal and intramodal information for each modality, and then utilizes a multimodal enhanced SSM for modality cohesion. 
Extensive experimental results on two large-scale depression datasets demonstrate the superior performance of our DepMamba over existing state-of-the-art methods.
Code is available at \url{https://github.com/Jiaxin-Ye/DepMamba}.


\end{abstract}



%% file: Section/1_intro.tex
\section{Introduction}
Depression is one of the most prevalent mental disorders, manifesting through a wide range of physiological symptoms such as weight loss and insomnia, with severe cases potentially leading to suicide or substance abuse~\cite{beck2009depression}.
Detecting depression faces two major challenges: (1) a steadily growing patient population, 
and (2) the high cost of manual diagnosis.
Consequently, there is an urgent need to develop an efficient depression detection system.

In recent years, multimodal-based methods, which integrate information from audio, video, and text modalities, have shown promising results in depression detection. These methods primarily focus on multimodal fusion, which can be categorized into three types: feature-level, decision-level, and model-level fusions. 
Feature-level fusion involves concatenating multiple modalities to learn a unified representation for depression detection~\cite{DBLP:conf/acii/HeJS15,DBLP:journals/inffus/CaiQLZHH20,DBLP:journals/tce/NingHYQTW24,DBLP:conf/mm/GuptaMXGSBPN14,DBLP:conf/mm/CumminsJDSGE13}. For example, Cai~\etal~\cite{DBLP:journals/inffus/CaiQLZHH20} introduce a linear combination technique to construct global representation from electroencephalogram signals of each modality. 
Decision-level fusion ensembles decision outputs from each modality to make a final classification~\cite{DBLP:conf/mm/NasirJSCG16,DBLP:journals/titb/ZhangSDLWH19,DBLP:journals/titb/ZhangCSLL22,AVEC2016,AVEC2019,DBLP:conf/iscc/RasipuramBMSS22}. Zhang~\etal~\cite{DBLP:journals/titb/ZhangCSLL22} propose a decision-level multimodal fusion based on multi-agent. 
Model-level fusion, considered the most efficient, learns the interrelationships among modalities~\cite{dvlog,DBLP:journals/inffus/FanZXFZZY24,DBLP:journals/tkde/TaoYLWH24}. For instance, Fan~\etal~\cite{DBLP:journals/inffus/FanZXFZZY24} utilize convolutional neural networks (CNN) to extract high-level unimodal features, while employing the Transformer model to enhance multimodal features. These methods demonstrate that multimodal cues can significantly improve the performance of depression detection.

However, existing methods still face two significant limitations: 
(1) \emph{inefficient long-range temporal modeling}, as demonstrated by CNNs being restricted by limited receptive fields, recurrent neural networks (RNNs) suffering from vanishing gradient issues, and self-attention mechanisms like Transformers grappling with computational inefficiency; 
and (2) \emph{sub-optimal multimodal fusion}, as exemplified by current techniques that tend to concentrate on learning either modal-shared or modal-specific features but are inadequate at concurrently capturing shared features while retaining modality-specific information.


To address these limitations, we propose an audio-visual progressive fusion Mamba model for efficient multimodal depression detection,  termed \textbf{\method}. Specifically, the DepMamba features two core designs: hierarchical contextual modeling and progressive multimodal fusion. 
First, we introduce CNN and Mamba blocks to extract features from local to global scales, enriching  contextual representation within long-range sequences. 
Second, we propose a multimodal collaborative State Space Model (SSM) that extracts intermodal and intramodal information for each modality by sharing state transition matrices. A multimodal enhanced SSM is then employed to process concatenated audio-visual features for improved modality cohesion. 
Extensive experimental results on two large-scale depression datasets demonstrate that \method outperforms existing state-of-the-art models in both accuracy and efficiency for depression detection.

\begin{figure*}[t]
    \centering
    \includegraphics[width=1\linewidth]{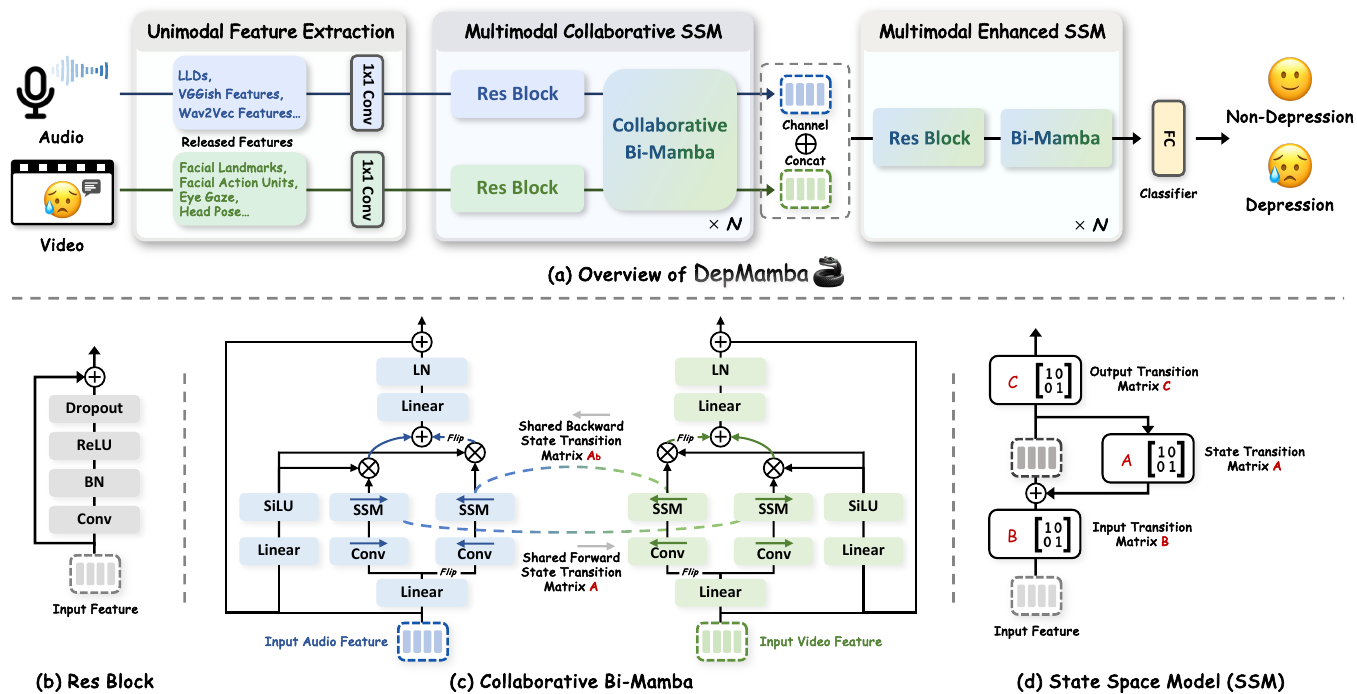}
    \caption{Overview of \method. For each modality, we first use a convolution layer to transform the released features of the dataset, which are then fed into multimodal collaborative and enhanced SSMs to extract complementary multimodal representations. Finally, a fully-connected layer receives the features and predicts the depression label. We utilize different colors to indicate the flow direction of different modalities. }
    \label{fig:main}
\end{figure*}

\noindent\textbf{Contributions.}\quad The contributions of this work are summarized as follows.
First, we introduce \method, a novel and efficient method that incorporates hierarchical modeling and progressive fusion, marking the first attempt of Mamba to depression detection. 
Second, we develop a hierarchical modeling combining CNN and Mamba for better learn local and global contextual representation. 
Third, we propose a progressive fusion with a key collaborative SSM that enhances intermodal fusion while preserving intramodal features, offering a new perspective on multimodal fusion. 
Finally, extensive experimental results demonstrate the superiority and efficiency of the proposed method compared to state-of-the-art baselines.



%% file: Section/2_method.tex
\section{Proposed Method}

\subsection{Preliminary of Mamba}

In recent years, the state space model (SSM) has developed rapidly~\cite{s4,s4d,h3,mamba1,mamba2,mamba3,DBLP:conf/kdd/BehrouzH24}, originating from classic control theory and offering linear scalability for long-range dependency modeling. The SSM introduces a hidden state $\boldsymbol{h}\left(t\right)\in\mathbb{R}^{N}$ to map the input $\boldsymbol{x}\left(t\right)\in\mathbb{R}^{L}$ to obtain the output $\boldsymbol{y}\left(t\right)\in\mathbb{R}^{L}$, where $N$ and $L$ denote the number of hidden states and sequence length. The continuous SSM system can be formulated as:
\begin{equation}
\label{eq1}
\begin{aligned}
    \boldsymbol{h'}(t) = \mathbf{A}\boldsymbol{h}(t) + \mathbf{B}\boldsymbol{x}(t), \,\,\,
    \boldsymbol{y}(t) = \mathbf{C}\boldsymbol{h}(t). 
\end{aligned}
\end{equation} 
where the state matrix $\mathbf{A}\in \mathbb{R}^{N\times N}$ and input/output projection matrixes $\mathbf{B}\in \mathbb{R} ^{N\times 1}$,$\mathbf{C}\in \mathbb{R} ^{N\times 1}$. The Mamba~\cite{mamba} further uses a time scale parameter $\Delta$ to discretize the continuous parameters $\mathbf{A, B}$ into $\mathbf{\overline{A}, \overline{B}}$, the zero-order hold (ZOH) principle is adopted by default. The discretized state-space equation is: $\mathbf{\overline{A}} = \exp \left( \mathbf{\Delta A}\right)$ and $\mathbf{\overline{B}} = \left( \mathbf{\Delta A}\right) ^{-1}\left( \exp \left( \mathbf{\Delta A}\right) -\mathbf{I}\right) \cdot \mathbf{\Delta B}$. After the discretization, the discretized version of Eq.~\eqref{eq1} with step size $\Delta$ can be rewritten in the following recurrent form:
\begin{equation}
\label{eq3}
\begin{aligned}
    \boldsymbol{h}_{t} = \mathbf{\overline{A}}\boldsymbol{h}_{t-1}+\mathbf{\overline{B}}\boldsymbol{x}_{t}, \,\,\,
    \boldsymbol{y}_{t} = \mathbf{C}\boldsymbol{h}_{t}.
\end{aligned}
\end{equation}
Finally, the Eq.~\eqref{eq3} can be equivalently transformed into the convolution form: $\mathbf{\overline{K}} =  ( \mathbf{C\overline{B},\ldots,C\overline{A}^{\mathrm{L}-1}\overline{B}}), \boldsymbol{y} = \boldsymbol{x}\circledast\mathbf{\overline{K}}$, where $\circledast$ denotes convolution operation and the global convolution kernel $\mathbf{\overline{K}}\in \mathbb{R} ^{L}$. Mamba has greatly advanced deep learning via its data-dependent mechanisms and efficiency. In this paper, we introduce bi-directional Mamba (Bi-Mamba)~\cite{vim} as the baseline, which comprehensively models long-range context. 

\subsection{Overview of \method}

As shown in Fig.~\ref{fig:main}(a), the proposed \method achieves hierarchical contextual modeling and progressive multimodal fusion comprising three vital components. 
(i) \textit{Unimodal feature extraction} first utilizes the released unimodal features for each modality as existing depression datasets~\cite{dvlog,lmvd} typically do not include raw signals due to privacy concerns. Then, these released features are respectively transformed into the same dimensional space through the 1D convolution.
(ii) \textit{Multimodal collaborative SSM (CoSSM)} aims to model hierarchical contextual information and aggregate modality-specific and modality-shared representations. Each CoSSM layer integrates two residual blocks (Res Blocks)~\cite{resnet} and a collaborative Bi-Mamba. 
(iii) \textit{Multimodal enhanced SSM (EnSSM)} first concatenates audio and visual features, and also models hierarchical information while enhancing multimodal cohesion with a ResBlock and a Bi-Mamba of each layer.
Finally, a linear layer is employed to perform depression detection. We introduce each part in the following Sections.
    

\input{Table/SOTA}

\subsection{Hierarchical Contextual Modeling}
\label{HCM}
For the CoSSM and EnSSM in Fig.~\ref{fig:main}, we propose to utilize CNN and Bi-Mamba to achieve hierarchical contextual modeling from local to global scales, effectively capturing the long-sequence audiovisual content.  
Specifically, the residual block with the small convolution kernel excels at capturing local temporal information. The Bi-Mmaba incorporates the two bidirectional SSMs for data-dependent global contextual modeling, which builds a bidirectional core memory of the context with selective attention mechanisms. 
Subsequently, these local and global features synergize to enrich the contextual representation. This process effectively extracts the inherent complementary features within long-range sequences, thereby enhancing overall performance.

\subsection{Progressive Multimodal Fusion}
Based on the modal-agnostic contextual modeling in Section~\ref{HCM}, we further introduce a two-stage progressive multimodal fusion. In the first stage (\ie CoSSM), for extracting complementary information between audio-visual modalities, we propose the collaborative Bi-Mamba to facilitate modality interaction. Specifically, the forward SSM and backward SSM of Bi-Mamba include 6 key parameters: forward matrices $\mathbf{\overline{A}},\mathbf{\overline{B}}, \mathbf{C}$ and backward matrices $\mathbf{\overline{A}}_\text{b},\mathbf{\overline{B}}_\text{b},\mathbf{C}_\text{b}$.  
The state transition matrices $\mathbf{\overline{A}},\mathbf{\overline{A}}_\text{b}$ have the most significant impact on the system, as they govern the evolution of the current hidden state, while $\mathbf{\overline{B}},\mathbf{\overline{B}}_\text{b}$ and $\mathbf{C},\mathbf{C}_\text{b}$ primarily influence the input and output states. 
Therefore, as shown in Fig.~\ref{fig:main}(c), we propose to share the bidirectional state transition matrices $\mathbf{A}$ and $\mathbf{A}_\text{b}$ across modalities, learning the contextual information shared between modalities. In contrast, $\mathbf{\overline{B}},\mathbf{\overline{B}}_\text{b}$ and $\mathbf{C},\mathbf{C}_\text{b}$ for different modalities remain independent to capture modality-specific information. The forward collaborative SSM can be formulated as:
\begin{equation}
\label{eq:cossm}
\begin{aligned}
    \boldsymbol{h}_{t}^\text{a} &= {\text{ }\mathbf{\overline{A}}\boldsymbol{h}_{t-1}^\text{a}}\,+{\,\,\,\mathbf{\overline{B^\text{a}}}\boldsymbol{x}_{t}^\text{a}}, 
    &\boldsymbol{y}_{t}^\text{a} = \mathbf{C}^\text{a}\boldsymbol{h}_{t}^\text{a},\\
     \boldsymbol{h}_{t}^\text{v} &= \underbrace{\mathbf{\overline{A}}\boldsymbol{h}_{t-1}^\text{v}}_{\text{Intermodal}}+\underbrace{\mathbf{\overline{B^\text{v}}}\boldsymbol{x}_{t}^\text{v},}_{\text{Intramodal}}
    &\boldsymbol{y}_{t}^\text{v} = \mathbf{C}^\text{v}\boldsymbol{h}_{t}^\text{v},
\end{aligned}
\end{equation} 
where $\boldsymbol{x}_{t}^\text{a},\boldsymbol{x}_{t}^\text{v}$, $\boldsymbol{y}_{t}^\text{a},\boldsymbol{y}_{t}^\text{v}$, and  $\boldsymbol{h}_{t}^\text{a},\boldsymbol{h}_{t}^\text{v}$ denote audio and visual input, output, and hidden features. $\mathbf{\overline{A}}$ is the shared forward state transition matrix, $\mathbf{\overline{B^\text{a}}},\mathbf{\overline{B^\text{v}}}$ and $\mathbf{C}^\text{a},\mathbf{C}^\text{v}$ are the input and output matrix for both modalities. Without introducing extra parameters, the CoSSM based on the control systems theory explicitly models intermodal shared and intramodal specific information, for complementary multimodal representation learning. 

Furthermore, in the second stage (\ie EnSSM), we first concatenate audio and visual output features from the CoSSM, and also utilize the Res-Block and Bi-Mamba to enhance multimodal cohesion with hierarchical contextual modeling. By employing a two-stage pipeline, we comprehensively integrate both intramodal and intermodal information, facilitating a more effective multimodal fusion.
Finally, we simply use a fully-connected layer for depression classification.

%% file: Table/SOTA.tex
\begin{table*}[t]
    \centering
    \caption{Performance comparison between the baseline models and our proposed \method on the D-vlog and LMVD datasets. Each model is run three times to obtain the results (mean \% $\pm$ stv \%). (Bold is the best)}
    \label{tab:SOTA}

    \begin{tabular}{lcccccccccc}
        \toprule
        \multirow{2}{*}{\textbf{Methods}} &\multicolumn{5}{c}{\textbf{D-Vlog}} & \multicolumn{5}{c}{\textbf{LMVD}}\\
        \cmidrule(r){2-6} \cmidrule(l){7-11}
         & \textbf{Accuracy} & \textbf{Precision} & \textbf{Recall} & \textbf{F1} & \textbf{Avg}  & \textbf{Accuracy} & \textbf{Precision} & \textbf{Recall} & \textbf{F1} & \textbf{Avg}  \\
        
        \midrule
        KNN~\cite{KNN} & 60.38\std{$\pm$1.42} & 61.41\std{$\pm$1.32} & 85.64\std{$\pm$4.48} & 71.47\std{$\pm$1.26} & 69.72 & 56.83\std{$\pm$1.64} & 57.43\std{$\pm$1.52} & 50.92\std{$\pm$4.44} & 53.92\std{$\pm$2.97} & 54.77 \\
        Bi-LSTM~\cite{bilstm} & 64.47\std{$\pm$2.60} & 67.68\std{$\pm$4.34} & 75.34\std{$\pm$4.97} & 71.10\std{$\pm$0.36} & 69.65 & 66.85\std{$\pm$1.92} & 65.81\std{$\pm$3.58} & 70.33\std{$\pm$4.79} & 67.83\std{$\pm$0.70} & 67.71\\
        TBN~\cite{TBN} & 63.21\std{$\pm$0.47} & 69.99\std{$\pm$4.28} & 65.31\std{$\pm$8.54} & 67.16\std{$\pm$2.61} &   66.42  &67.94\std{$\pm$1.76} &  67.06\std{$\pm$2.92} & 70.33\std{$\pm$3.96} & 68.56\std{$\pm$1.35}& 68.47 \\
        STFT~\cite{STFT} & 61.79\std{$\pm$0.94} & 64.67\std{$\pm$3.34} & 77.51\std{$\pm$15.26} & 69.77\std{$\pm$4.77} & 68.44 & 67.76\std{$\pm$1.09} & 69.20\std{$\pm$2.65} & 64.01\std{$\pm$7.20} & 66.23\std{$\pm$3.03} & 66.80 \\
        TFN~\cite{TFN} & 67.14\std{$\pm$1.09} & \best{72.38\std{$\pm$1.22}} & 70.19\std{$\pm$3.08} & 71.23\std{$\pm$1.43} & 70.23 & 63.93\std{$\pm$1.89} & 64.08\std{$\pm$2.25} & 62.64\std{$\pm$1.10} & 63.34\std{$\pm$1.59}  & 63.50\\
        DepTrans~\cite{dvlog} & 62.89\std{$\pm$4.23} & 64.43\std{$\pm$5.58} & 84.82\std{$\pm$13.37} &72.54\std{$\pm$1.09} & 71.17  & 61.93\std{$\pm$3.01} & 60.36\std{$\pm$4.19} & 72.16\std{$\pm$13.43} & 65.08\std{$\pm$3.52}& 64.88 \\
        TAMFN~\cite{TAMFN} & 67.45\std{$\pm$2.45} & 68.08\std{$\pm$2.50} & 82.93\std{$\pm$0.81} & 74.75\std{$\pm$1.25} & 73.30 & 70.49\std{$\pm$0.55} & \best{71.15\std{$\pm$2.73}} & 68.86\std{$\pm$5.19} & 69.84\std{$\pm$1.30} & 70.09\\
        \midrule
        \textbf{\method} &\best{68.87\std{$\pm$1.89}} & 68.19\std{$\pm$1.85} & \best{86.99\std{$\pm$0.81}}  & \best{76.44\std{$\pm$1.02}} & \best{75.12} & \best{72.13\std{$\pm$1.09}} & 70.18\std{$\pm$1.66} & \best{76.56\std{$\pm$2.54}} & \best{73.20\std{$\pm$1.01}}& \best{73.02}  \\
        \bottomrule
    \end{tabular}
\end{table*}

%% file: Section/3_exp.tex
\section{Experiments}

\subsection{Experimental Setup}
\noindent{\textbf{Datasets.}}\quad
To demonstrate the effectiveness of the proposed \method with State-Of-The-Art (SOTA) methods, we conduct extensive experiments on Depression Vlog (D-Vlog)~\cite{dvlog} and Large-Scale Multimodal Vlog Dataset (LMVD)~\cite{lmvd}, which are currently large-scale depression datasets. The D-Vlog and LMVD datasets are both audio-visual depression detection datasets for real-world scenarios, which are collected from various vlog videos published on social media platforms such as YouTube and TikTok. To protect user privacy, both datasets utilize the dlib~\cite{dlib} toolkit to extract 68 facial landmarks as visual features. Additionally, the D-Vlog employs the OpenSMILE~\cite{opensmile} toolkit to extract 25 low-level descriptors, while the LMVD extracts VGGish~\cite{vggish} features as audio features, respectively.
The D-Vlog is split into train, validation and test sets with a 7:1:2 ratio, and we split the LMVD with a 8:1:1 ratio. In our experiments, we run each model three times to avoid randomness.

\noindent{\textbf{Implementation details.}}\quad
For \method, we train the model using the Adam optimizer~\cite{adam/KingmaB14} with a learning rate of 1e-4, batch size 16, and a 24GB NVIDIA RTX 3090 GPU. The binary cross-entropy criterion is used as the objective function and the overall epoch is set to 120. The convolutions of unimodal feature extraction are employed with 1 kernel size and 256 channel size. We use a hidden dimension size of 256, a state dimension of 12 for the D-Vlog and 16 for the LMVD, an expansion coefficient of 4 as the configuration of each Bi-Mamba block, and a layer number of 1 for both CoSSM and EnSSM to train our \method. 

\noindent{\textbf{Evaluation metrics.}}\quad
Following previous works~\cite{dvlog,lmvd,TAMFN}, we introduce four widely-used metrics to evaluate performance including Accuracy, Precision, Recall (also known as sensitivity), and F1-score. 
These matrices can evaluate overall performance on both balance and imbalance scenarios. 

\begin{figure*}[ht]
    \centering
    \includegraphics[width=1.\linewidth]{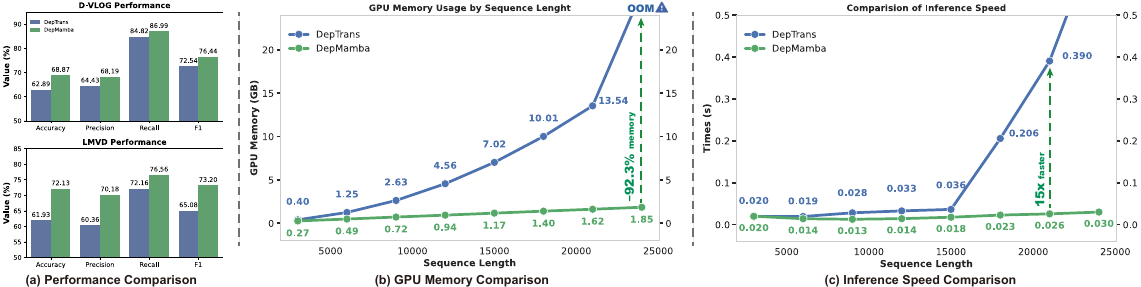}
    \caption{Performance and efficiency comparisons between Transformer-based DepTrans and our Mamba-based DepMamba.}
    \label{fig:eff}
\end{figure*}

\subsection{Comparison Results}
To demonstrate the effectiveness of our \method, we compare several representative multimodal depression detection methods on two large-scale datasets. 
As shown in Table~\ref{tab:SOTA}, we utilize 4 metrics and their average value to evaluate these methods. It can be observed that \method shows superior performance on both datasets with average values of 75.12 and 73.02, gaining 1.82\% and 2.93\% improvements with the second-best method. 
In terms of recall, \method achieves the best performance, indicating that it is more effective at identifying individuals with depression, thus minimizing the risk of false negatives in real-world scenarios. 
Additionally, compared to multimodal fusion methods such as TBN~\cite{TBN}, TFN~\cite{TFN}, and TAMFN~\cite{TAMFN}, the proposed CoSSM and EnSSM enhance \method's ability to capture complementary representations of depression in audio-visual data, thereby improving the performance in depression detection.

\input{Table/Ablation}
\subsection{Ablation Study}
We conduct ablation studies to validate the effectiveness of several design components: (i) \emph{modality selection} for exploring the impacts of modality (\ie single modality or multi-modality), (ii) \emph{hierarchical modeling} for analyzing the effects of local (\ie CNN) and global (\ie Mamba) temporal modeling, and (iii) \emph{progressive modeling} for evaluating the influences of collaborative and enhanced fusions.
The average results of 4 metrics on two datasets are shown in Table~\ref{tab:ablation}. We have the following observations. 
\emph{First}, multimodal data provides more comprehensive and effective information compared to unimodal one, with 2.81\% and 3.60\% improvements on the two datasets.
\emph{Second}, global temporal modeling significantly boosts depression detection more than local modeling, as patients often pause while speaking leading to long-range temporal content. The removal of Mamba leads to performance drops of 4.27\% and 1.70\%, whereas excluding CNN results in only 0.50\% and 0.42\% reductions. 
\emph{Finally}, progressive fusion effectively facilitates the interaction between audio-visual information, and extracts intra-modal and inter-modal knowledge to reach better performance.


\subsection{Efficiency Analysis}
As shown in Fig.~\ref{fig:eff}, we evaluate the efficiency of the Mamba-based \method compared to the Transformer-based DepTrans. Results show that the proposed \method outperforms DepTrans on both datasets with the 3.95\% and 8.14\% average improvements. \method is also more computation and memory efficient than DepTrans in dealing with long-range sequence modeling. For example, across 8 different sequence lengths, \method is $15\times$ faster than DepTrans with a length of 21000 and saves 92.3\% GPU memory with a length of 24,000. 
While the GPU memory usage of the Transformer-based method increases exponentially with increasing length~\cite{DBLP:conf/nips/VaswaniSPUJGKP17}, \method exhibits linear growth with obvious advantages in the practical application. 

%% file: Table/Ablation.tex
\begin{table}[t]
    \centering
    \caption{Ablation study. ‘w/o’ denotes removing the component.}
    \label{tab:ablation}

    \begin{tabular}{lcc}
        
        \toprule
        \textbf{Methods}& \textbf{D-Vlog} & \textbf{LMVD} \\
        \midrule
        \multicolumn{3}{l}{\textbf{\textit{Modality Selection}}} \\
        only Audio & 71.96\std{$\pm$1.39} & 62.35\std{$\pm$4.27}  \\
        only Visual & 72.31\std{$\pm$2.93} & 69.42\std{$\pm$4.25}  \\
        \specialrule{0em}{1.5pt}{1.5pt}

        \multicolumn{3}{l}{\textbf{\textit{Hierarchical Modeling}}} \\
        wo CNN & 74.62\std{$\pm$0.26} & 72.60\std{$\pm$3.06}  \\
        wo Mamba & 70.85\std{$\pm$2.01} & 71.32\std{$\pm$0.61}  \\
        \specialrule{0em}{1.5pt}{1.5pt}

        \multicolumn{3}{l}{\textbf{\textit{Progressive Modeling}}} \\
        wo CoSSM & 71.00\std{$\pm$0.98} & 70.65\std{$\pm$0.66}  \\
        wo EnSSM & 72.91\std{$\pm$2.64} & 71.15\std{$\pm$5.14}  \\
        
        \midrule
        \textbf{\method} &\best{75.12\std{$\pm$1.09}} & \best{73.02\std{$\pm$0.98}}   \\
        \bottomrule
    \end{tabular}
\end{table}

%% file: Section/4_summary.tex
\section{Conclusion}

We propose \method, the first multimodal depression detection method based on the Mamba to address the challenges of inefficient long-range temporal modeling and sub-optimal multimodal fusion, which can learn efficient representations through hierarchical modeling and progressive fusion, serving as a foundation for future research in multimodal fusion. 
The experimental results consistently indicate that the \method achieves superiority and efficiency and further with a lower number FLOPs of and faster inference speed than Transformer-based methods. Moreover, for detecting depression, we observe that multimodal features are more important than unimodal features and global temporal modeling is impactful more than local modeling. 
In the future, we will investigate the hybrid architecture based on the Mamba and Transformer for better representation generalization in cross-domain depression detection tasks.